\documentclass[a4paper,fleqn]{cas-sc}
\usepackage[sort&compress ,numbers]{natbib}
\usepackage{graphicx}
\usepackage{epsf}
\usepackage{rotating}
\usepackage{amsmath}
\usepackage{graphicx,epsfig}

\usepackage{dcolumn}
\usepackage{bm}
\usepackage{color}

\def\vmax{{v_{\rm max}}}

\def\vi{v_i}

\def\xit{x_i^t}
\def\yit{y_i^t}
\def\xito{x_i^{t+1}}
\def\yito{y_i^{t+1}}
\def\xjt{x_j^t}
\def\yjt{y_j^t}

\def\na{\bar{n}_{\rm arrival}}
\def\nac{\bar{n}^*_{{\rm arrival}_{\rm controlled}}}       

\def\ng{\bar{g}}
\def\git{g_i^t}
\def\gito{g_i^{t+1}}
\def\Pit{P_i^t}
\def\etait{\eta_i^t}
\def\Dxit{\Delta x_i^t}
\def\Dyit{\Delta y_i^t}
\def\Dg{\Delta g}
\def\ait{a_i^t}

\def\aito{a_i^{t+1}}

\newcommand{\sgn}[1]{{\rm Sign}\left( #1\right)}

\newcommand{\tuple}[2]{\Big(#1, #2\Big)}

\newcommand{\req}[1]{Eq.~(\ref{#1})}

\newcommand{\fig}[1]{Figure~\ref{#1}}

\newcommand{\cut}[1]{{}}

\newcommand{\red}{\textcolor{red}}

\def\tsc#1{\csdef{#1}{\textsc{\lowercase{#1}}\xspace}}
\tsc{WGM}
\tsc{QE}
\tsc{EP}
\tsc{PMS}
\tsc{BEC}
\tsc{DE}
\begin{document}
\let\WriteBookmarks\relax
\def\floatpagepagefraction{1}
\def\textpagefraction{.001}
\shorttitle{Adaptive strategies for route selection}
\shortauthors{T. S. Tai and C. H. Yeung}

\title [mode = title]{Adaptive strategies for route selection en-route in transportation networks}      

\author%
[1]
{Tak Shing Tai}

\author%
[1]
{Chi Ho Yeung}
\cortext[cor1]{Corresponding author}
\ead{chyeung@eduhk.hk}

\address[1]{Department of Science and Environmental Studies, The Education University of Hong Kong, Tai Po, Hong Kong}

\begin{abstract}
We examine adaptive strategies adopted by vehicles for route selection en-route in transportation networks. By studying a model of two-dimensional cellular automata, we model vehicles characterized by a parameter called path-greediness, which corresponds to the tendency for them to travel to their destinations via the shortest path. The path-greediness of each individual vehicle is updated based on the local traffic conditions, to either keep the vehicle travels via a shorter path in an un-congested region or to explore longer diverted paths in a congested region. We found that the optimal number of steps to trigger an update of path-greediness is dependent on the density of vehicles, and the magnitude of path-greediness increment affects the macroscopic traffic conditions of the system. To better coordinate vehicles in denser networks, the update on the tendency for vehicles to travel via the shorter paths should be gradual and less frequent.
\end{abstract}

\maketitle
\section{Introduction}

Heavy traffic in urban area causes severe traffic congestion. The congestion reduces social and economic outputs because they increase the time cost and the fuel cost of drivers. To reduce the economic loss, different studies have suggested various solutions such as the coordination of users' routes \cite{WATLING2018254,natCom16,Yeung13717} to reduce traffic congestion. These methods, in general, aim to distribute traffic more evenly in the network and to reduce the traffic flow on major routes. These relations between the interaction among vehicles and the macroscopic phase of the transportation networks is discussed in different studies \cite{PhysRevE.84.046110, SIQUEIRA20161}. Since it is hard to study the interaction between vehicles onsite, mathematical models and computer simulations are often employed. For example, cellular automaton models suggested by Nagel and Schreckenberg \cite{NSModel} are often employed to study vehicular dynamics and to characterize the macroscopic behavior of the networks \cite{PhysRevE.83.061150}. In some studies, we see that the vehicles traveled by the suggested routes after coordination can optimize the traffic flow in the network. 

However, even vehicular routes are coordinated and recommended to road users, traffic conditions may be affected by road accidents or drivers who do not follow the suggested routes. In reality, drivers respond to the changes in traffic conditions en-route, and to decide whether the currently selected route is beneficial to them. For instance, when drivers think that their current routes are no longer optimal, they may switch to a new route. However, drivers usually do not have information on the global traffic conditions inside their vehicles so they are lack of the sufficient information to optimize their routes. Therefore, drivers usually make decisions based on local information, and the worst case is that they may switch to a new route with more serious traffic jam. Without the global traffic information, the chances of making wrong decision on route selection increase, thus worsening traffic congestion.

Since traffic conditions change from time to time, it is important to study the dynamics of transportation networks and to derive algorithms adaptive to the dynamically changing conditions to optimize the networks \cite{PhysRevE.83.016102,FRIESZ2019309}. For instance, the dynamical data of vehicle speed are collected and analyzed to optimize traffic flow in real-time~\cite{PhysRevE.83.047101,HADDAD201317}. Other studies suggest that traffic flow can be regulated by controling traffic light signals to coordinate vehicles in the area of severe traffic congestion \cite{LEE20171,YU2018302}; some studies show that adaptive signals are better than non-adaptive ones in both free flow state and congested state, and some adaptive algorithms can reduce the total traveling time by around $50\%$ \cite{LEE20171,LEE2017376,MOHEBIFARD2019252}. There are also studies which optimize traffic by considering dynamical trajectories of vehicles and formulate an objective function with different traffic parameters such as vehicle flows, queue lengthand other important factors including sacrificing minor streams \cite{YAN2019266,YAO2019211,CHACOMA2021125763}. These studies found that different states of traffic can be observed depending on the density and movement of vehicles \cite{LI2015349,REGRAGUI20181273}.

In this study, we will reveal the impact of the change in the tendency of vehicles to travel via the shortest path, when they are en-route to destinations. By studying a model of two-dimensional cellular automata \cite{PhysRevE.100.012311}, we model vehicles characterized by a parameter called \emph{path-greediness}, which corresponds to the tendency for them to travel to their destination via the shortest path. Unlike our previous work \cite{PhysRevE.100.012311} which mainly focused on \emph{non-adaptive} routing, here we show the effectiveness of routing protocols \emph{adaptive} to local traffic information. We will examine the impact of various update strategies of path-greediness on the transportation networks in terms of performance metrics such as the total number of arrivals. Specifically, despite all strategies to be studied are adaptive, we found that the network performance varies greatly depending on both the  frequency and the magnitude of updates . In addition, the optimal strategy on networks with different density of vehicles are different. The efficacy of adaptive routing strategies has only been briefly introduced in~\cite{PhysRevE.100.012311}, but will be comprehensively examined in the present study.

The rest of this paper is organized as follows: Section \ref{sec_model} introduces our two-dimensional cellular automata model of transportation networks. Section \ref{sec_results} shows the results of comparison and evaluation of different adaptive routing strategies. Finally, the conclusion is given in Section \ref{sec_conclusion}.

\section{Model}
\label{sec_model}
We consider a model of transportation networks on a two-dimensional $L \times L$ square lattice with periodic boundary condition, where each lattice site is labeled by a coordinate $(x,y)$, with $x,y=1, ..., L$. There are $N$ vehicles, labeled by $i=1, ..., N$ in the network and they run on the network by hoping between neighboring sites. Each site can only be occupied by one vehicle. We denote the density of vehicles on the network to be $\rho=N/L^2$. Initially, for each vehicle $i$, random origin and destination $(X_i,Y_i)$ are drawn. The vehicle travels from their origin to their destination. When a vehicle arrives at its destination, a new site on the network is drawn as the new destination of the vehicle so the vehicle travels continuously. 

We denote the coordinate of vehicle $i$ at time $t$ to be $(\xit,\yit)$ and its speed to be $\vi$. For simplicity, we consider the speed limit $\vmax$ to be 1, such that vehicles can only move to a neighboring site in each time step. Each vehicle $i$ is picked to move to a new coordinate $(\xito,\yito)$ according to a \emph{routing strategy}. The simulation repeats for $T$ steps, and various quantities of interest in the system such as number of arrivals are measured.

\subsection{Probabilistic Routing}
In this subsection, we will describe the probabilistic routing strategy introduced in ~\cite{PhysRevE.100.012311}, which we will refer to as the \emph{non-adaptive} or \emph{controlled} scenarios as the algorithmic parameters are not adaptive to the change in traffic condition. For each vehicle $i$ with coordinate $(\xit,\yit)$ at time $t$, the coordinate is updated by a probabilistic routing protocol \cite{PhysRevE.100.012311}, given by
\begin{align}
(\xito,\yito)=(\xit,\yit)+\etait (\Dxit,\Dyit)
\end{align}
where $(\Dxit,\Dyit)$ represents the intended movement of vehicle $i$. A variable $\etait=1$ corresponds the case that the site of the next intended movement of vehicle $i$ is empty at time $t$ and the intended movement is valid. Otherwise, the intended movement is invalid and $i$ should stay in the site $(\xit,\yit)$. In other words, $\etait$ is given by
\begin{align}
\etait =
\begin{cases}
0, &\hspace{-0.2cm}\mbox{if $\tuple{\xit}{\yit}\!+\!\tuple{\Dxit}{\Dyit}\!=\!\tuple{\xjt}{\yjt}, \exists j$}
\\
1, &\hspace{-0.2cm}\mbox{otherwise}
\end{cases}
\end{align}

\begin{figure}[h]
\centering
\begin{minipage}{18pc}
\includegraphics[width=18pc]{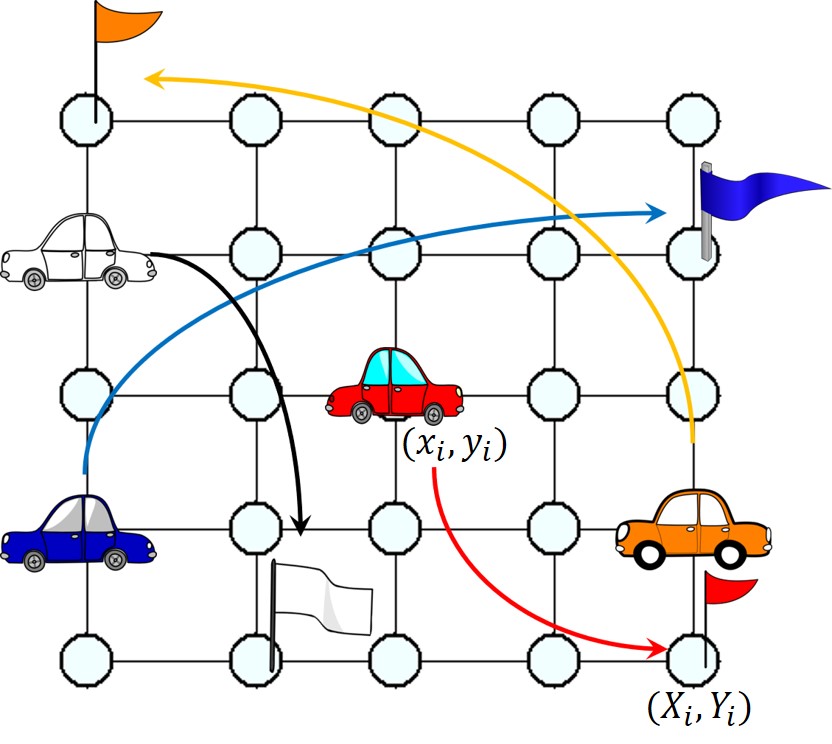}
\end{minipage}

\caption{An example of four vehicles and their destinations in a network with $5 \times 5$ square lattice}
\label{grid}
\end{figure}

Next, we introduce a parameter called \emph{path-greediness} $\git$, with $0\le\git\le 1$, which determines whether vehicle $i$ moves closer to their destination or randomly in the network at time $t$.  When $\git=1$, vehicle $i$ moves with the shortest path to its destination and when $\git=0$, vehicle $i$ move randomly. Specifically, when vehicle $i$ has not arrived at neither the $x$ nor $y$ coordinate of its destination (i.e., $\xit \neq X_i $ and $\yit \neq Y_i$), the intended movement $(\Dxit, \Dyit)$ of vehicle $i$ at time $t$ is 
\begin{align}
\label{eq_random1}
\left(\Delta x_i^t, \Delta y_i^t\right)=
\begin{cases}
\tuple{0}{\Delta \tilde{y}}, &\mbox{with a probability $\frac{1+\git}{4}$},
\\
\tuple{\Delta \tilde{x}}{0}, &\mbox{with a probability $\frac{1+\git}{4}$},
\\
\tuple{0}{-\Delta \tilde{y}}, &\mbox{with a probability $\frac{1-\git}{4}$},
\\
\tuple{-\Delta \tilde{x}}{0}, &\mbox{with a probability $\frac{1-\git}{4}$},
\\
\end{cases}
\nonumber\\
\end{align}
where $\Delta \tilde{y}$ and  $\Delta \tilde{x}$ are the greedy directions
\begin{align}
\label{eq_deltay}
\Delta \tilde{y}  = \sgn{Y_i-\yit} \sgn{\frac{L}{2}-|Y_i-\yit|}
\\
\label{eq_deltax}
\Delta \tilde{x}  = \sgn{X_i-\xit} \sgn{\frac{L}{2}-|X_i-\xit|}
\end{align}
with the sign function $\sgn{x}=1$ when $x \geq 0$, and otherwise $\sgn{x}=0$. Under periodic boundary condition, the vehicle moves in a  direction closer to the destination, so the first sign function determines the $x$ and $y$ direction from the present site to the destination and the second sign function reverses the movement direction when it is shorter to move across the boundary.

On the other hand, when vehicle $i$ has arrived at either the x or y coordinate of its destination (i.e., $\xit = X_i $ and $\yit \neq Y_i$, or $\xit \neq X_i $ and $\yit = Y_i)$, For example, if $\xit = X_i $, $(\Dxit, \Dyit)$ become
\begin{align}
\label{eq_random2}
\left(\Delta \xit, \Delta \yit\right)=
\begin{cases}
\tuple{0}{\Delta \tilde{y}},&\mbox{with a probability $\frac{1+3\git}{4}$},
\\
\tuple{1}{0},&\mbox{with a probability $\frac{1-\git}{4}$},
\\
\tuple{0}{-\Delta \tilde{y}},&\mbox{with a probability $\frac{1-\git}{4}$},
\\
\tuple{-1}{0}, &\mbox{with a probability $\frac{1-\git}{4}$}.
\\
\end{cases}
\end{align}
Similarly, if $\yit= Y_i$, then the $(\Dxit, \Dyit)$ become
\begin{align}
\label{eq_random3}
\left(\Delta \xit, \Delta \yit\right)=
\begin{cases}
\tuple{\Delta\tilde{x}}{0}, &\mbox{with a probability $\frac{1+3\git}{4}$},
\\
\tuple{0}{1}, &\mbox{with a probability $\frac{1-\git}{4}$},
\\
\tuple{-\Delta\tilde{x}}{0}, &\mbox{with a probability $\frac{1-\git}{4}$},
\\
\tuple{0}{-1}, &\mbox{with a probability $\frac{1-\git}{4}$},
\\
\end{cases}
\end{align}

\subsection{Strategies to update path-greediness}

In the paper, we study the mechanism by which path-greediness $\git$ of vehicle $i$ is updated according to the local traffic information of $i$, which we will refer to as the \emph{adaptive} strategies. We remark that such strategies are only briefly introduced in~\cite{PhysRevE.100.012311} without a mathematical formulation. There are two parameters that affect the update of path-greediness $\git$, namely (i) the number of steps $P$ of consecutive waiting or movements to trigger the update, and (ii) the increment or decrement magnitude $\Dg$ when $\git$ is updated. First, we denote $\Pit$ to be the number of consecutive movements of vehicle $i$ at time $t$ in recent $P$ steps, given by 
\begin{align}
\Pit = \sum_{t'=t-P+1}^{t} \eta_i^{t'}
\end{align}
When $\Pit=0$ or $\Pit=P$, it implies that vehicle $i$ is stuck or free-flowing for consecutive $P$ steps, respectively, which triggers an update of $\gito$. We further denote a variable $\aito$ given by 
\begin{align}
\aito =
\begin{cases}
1, &\hspace{-0.2cm}\mbox{if } \Pit=P
\\
-1, &\hspace{-0.2cm}\mbox{if } \Pit=0
\\
0, &\hspace{-0.2cm}\mbox{otherwise}
\end{cases}
\end{align}
such that the path-greediness of vehicle $i$ at time $t+1$ is updated by
\begin{align}
\label{eq_g_update}
\gito = 
\begin{cases}
0, &\mbox{if $\git + \ait\Dg<0$}
\\
1, &\mbox{if $\git + \ait\Dg>1$}
\\
\git + \ait\Dg, &\mbox{otherwise,}
\end{cases}
\end{align}
where the first two cases keep $0\le\gito\le 1$ as we have defined in \req{eq_random1}. According to Eq.~(\ref{eq_g_update}), path-greediness $\gito$ increases when vehicle $i$ has moved consecutively for $P$ steps, implying that the local traffic condition is good and vehicle $i$ can increase its tendency to travel to its destination via the shortest path. On the other hand, when vehicle $i$ is stuck for $P$ steps, it reduces its tendency to travel via the short path and to explore longer diverted paths to the destination. When all vehicles reduce their path-greediness, traffic congestion may be suppressed too.

\begin{table}[width=.9\linewidth,cols=3,pos=h]
\caption{Major model parameters.}\label{parameter}
\begin{tabular*}{\tblwidth}{@{} LLLL@{} }
\toprule
Description & Symbol & Value\\
\midrule
System size & $L \times L$ & $20 \times 20$\\
Simulation time & $T$ & $300000$ \\
Equilibrium time & $T_e$ & $250000$  \\
Path-greediness & $g$ & - \\
Magnitude of change of path-greediness & $\Dg$ & - \\
Number of consecutive waiting \red{or} movements before path-greediness is updated & $P$ & - \\
Average path-greediness & $\ng$ & - \\
Average arrival count & $\na$ & - \\
Magnitude of change of average path-greediness & $\Delta \ng$ & - \\
Magnitude of change of average arrival count & $\Delta \na$ & - \\
\bottomrule
\end{tabular*}
\end{table}

\subsection{Quantities of interest}

We conduct computer simulations on the model with over 1000 instances and assume that the simulations start at $t=0$ and equilibrate at $t=T_e$, before they end at $t=T$. The \emph{average arrival count} $\na$  , which corresponds to  the total number of  vehicle arrival at destinations per time step, will be measured to quantify the performance of the system. The larger the value of $\na$, the better the performance of the system. On the other hand, we will also measure the \emph{average path-greediness} $\ng$,
\begin{align}
\ng = \frac{1}{(T-T_e)N} \sum_{t=T_e}^{T} \sum_{i=1}^{N} \git
\end{align}
as path-greediness is self-organized by individual vehicles in the network.

\section{Results}
\label{sec_results}
We examine the dependence of the system performance  on  the parameters $\Dg$ and $P$. A higher value of $\Dg$ increases the magnitude of response to the traffic conditions while a higher value of $P$ increases the rate of response. If the value of $\Dg$ is too large, the system may result in traffic congestions since a large value of  $ng$ is more likely to lead to traffic congestion \cite{PhysRevE.100.012311}. Therefore, a large value of $\Dg$ may not benefit the system. Similarly, a large value of $P$ is not good for the system since vehicles may over-react to the traffic conditions. 

Here we examine different scenarios of the parameters, including  cases with $\Dg=1$, such that $\git$ of vehicle $i$ at time $t$ can only take two possible values, either $\git=0$ or $\git=1$, i.e. vehicle $i$ can either move totally randomly or with the shortest path, respectively. We start the simulations with the initial values of $g_i^0=0$ for all vehicle $i$; we observed that the results are independent of the initial condition of $\git$.

\begin{figure}[h]
\begin{minipage}{18pc}
\includegraphics[width=18pc]{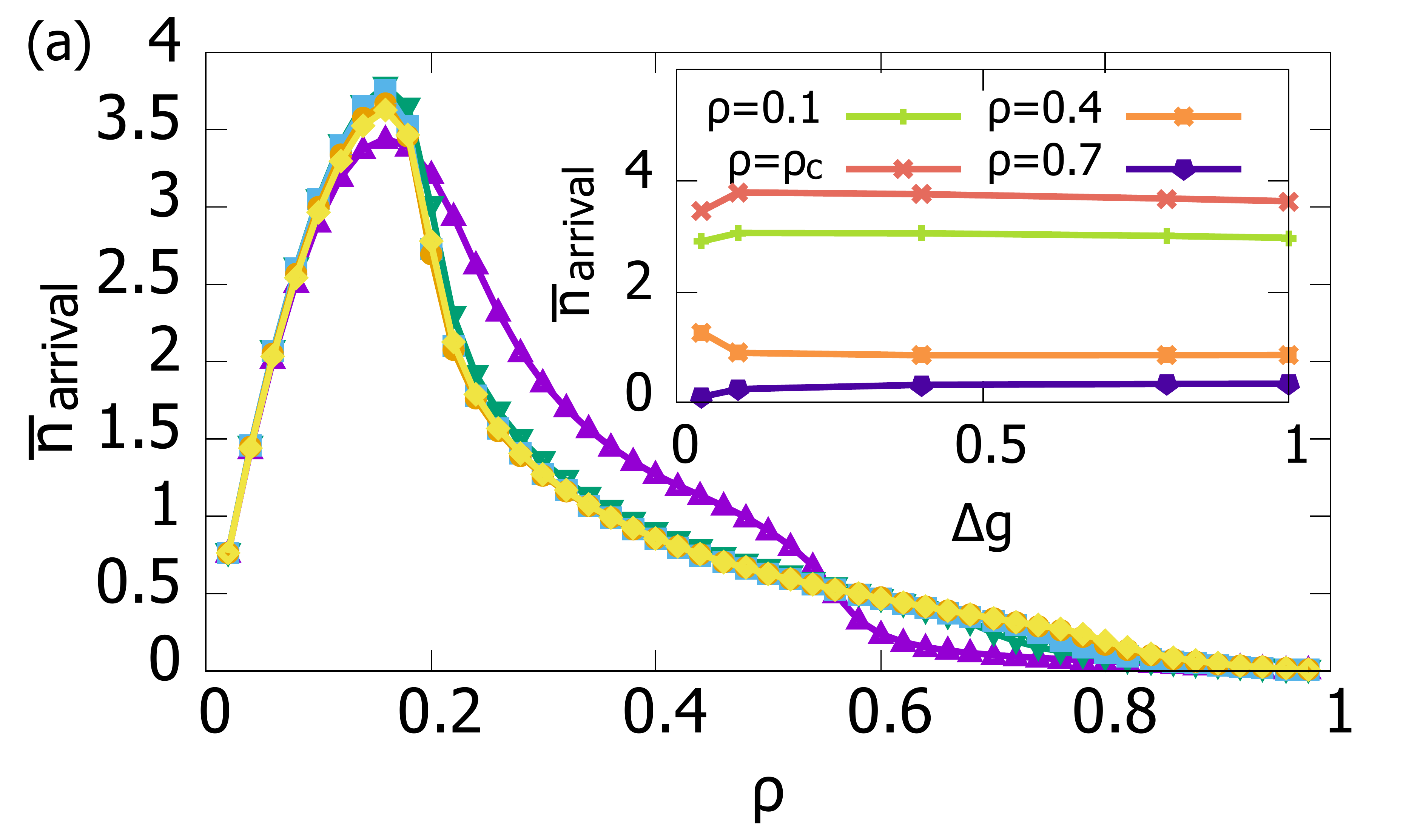}
\end{minipage}\hspace{2pc}%
\begin{minipage}{18pc}
\includegraphics[width=18pc]{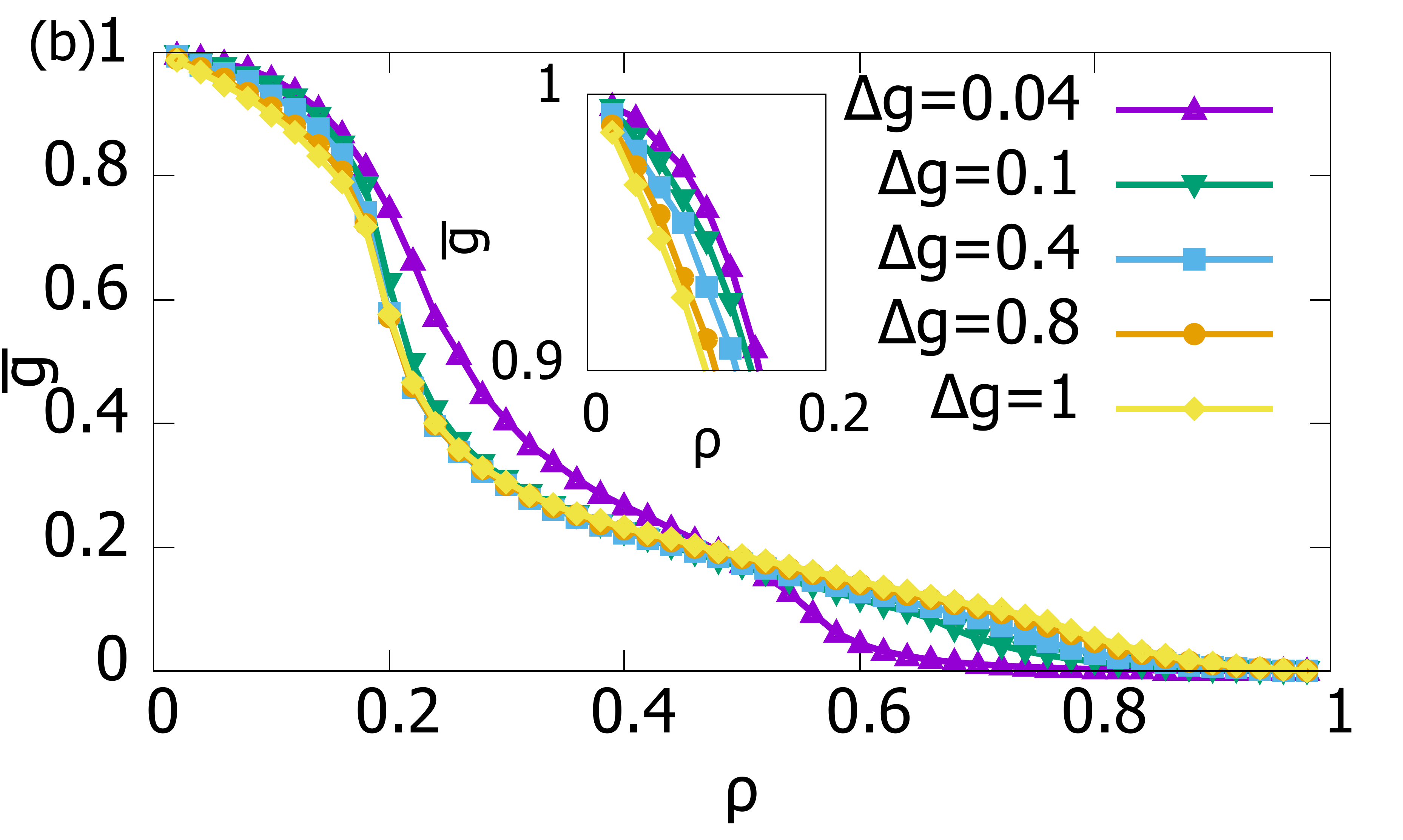}
\end{minipage}

\caption{The simulation results of (a) the average arrival count $\na$ and (b) the average path-greediness $\ng$ as a function of the density of vehicles $\rho$, where path-greediness is updated after $P=3$ consecutive waiting or movements. Results for various values of magnitude change of path-greediness $\Dg$ are obtained on square lattice with $L=20$ and periodic boundary condition. Inset: the simulation results of $\na$ as a function of $\Dg$ is shown in the inset of (a) and the enlarged plots of $\ng$ in the regime with small $\rho$ is shown in the inset of (b).}
\label{diffDg}
\end{figure}

As shown in \fig{diffDg}(a), we observed that the overall trend of the arrival count $\na$ is similar in cases with different values of $\Dg$ and $P=3$. We can roughly identify two states, namely the \emph{free-flow state} when $\na$ increases with vehicle density $\rho$ when $\rho \leq 0.2$, and the \emph{congested state} when $\na$ decreases with $\rho$ when $\rho>0.2$. The two states are defined similarly to those in the conventional fundamental diagram of transportation networks, where traffic flows increase and decrease respectively with vehicle density~\cite{maerivoit2005_physicsReports}. We denote $\rho_c$ to be the threshold density of vehicles at the maximum value of $\na$, when the system transits from the free-flow state to the congested state.

As for the system performance with different $\Dg$, in the free-flow state, cases with smaller $\Dg$ generally perform better, especially at $\rho=\rho_c$. We can see that in \fig{diffDg}(a) and its inset, the case with $\Dg=0.1$ attains the maximum value of $\na$ compared to cases with other values of $\Dg$. However, when $\Dg$ further decreases such that $\Dg=0.04$, $\na$ becomes smaller than all the other cases studied. Furthermore, for case with $\Dg=1$, which is the system with only two possible values of $\git$ i.e. $\git=0$ or $\git=1$, its maximum value of $\na$ is slightly smaller than that with $\Dg \geq 0.1$; it is because the case with $\Dg=1$ has less flexibility and vehicles in this case are often characterized with $\git=0$ and move randomly, while the vehicles in other cases with lower values of $\Dg$ have higher flexibility to adjust to an optimal positive $\git$. 

However, when vehicle density $\rho$ further increases to the congested state with $0.2< \rho\leq 0.56$, we observed that in \fig{diffDg}(a) and its inset, the case with $\Dg=0.04$ outperforms other cases in terms of $\na$, even though its values of  $\na$ are lower than those of the other cases in the free-flow state. We note that with a large value of $\Dg$ in the congested state, vehicles may not be able to have gradual and adaptive response to the traffic conditions, and they may over-react which lead to more severe congestion. We also observed that, in \fig{diffDg}(b), the value of average path-greediness $\ng$ in the case with $\Dg=0.04$ is higher than that in cases with $\Dg>0.04$, implying that vehicles in the case with $\Dg=0.04$ are able to self-adjust gradually to a higher path-greediness. These results show that the magnitude of responses is important in the congested state with $\rho\leq 0.56$ and a gradual change of $g$ is preferred. However, when $\rho > 0.56$, a larger value of $\Dg$ is preferred as $na$ for the case with $\Dg=0.04$ drops and its $\ng$ rises.

\begin{figure}[h]
\begin{minipage}{18pc}
\includegraphics[width=18pc]{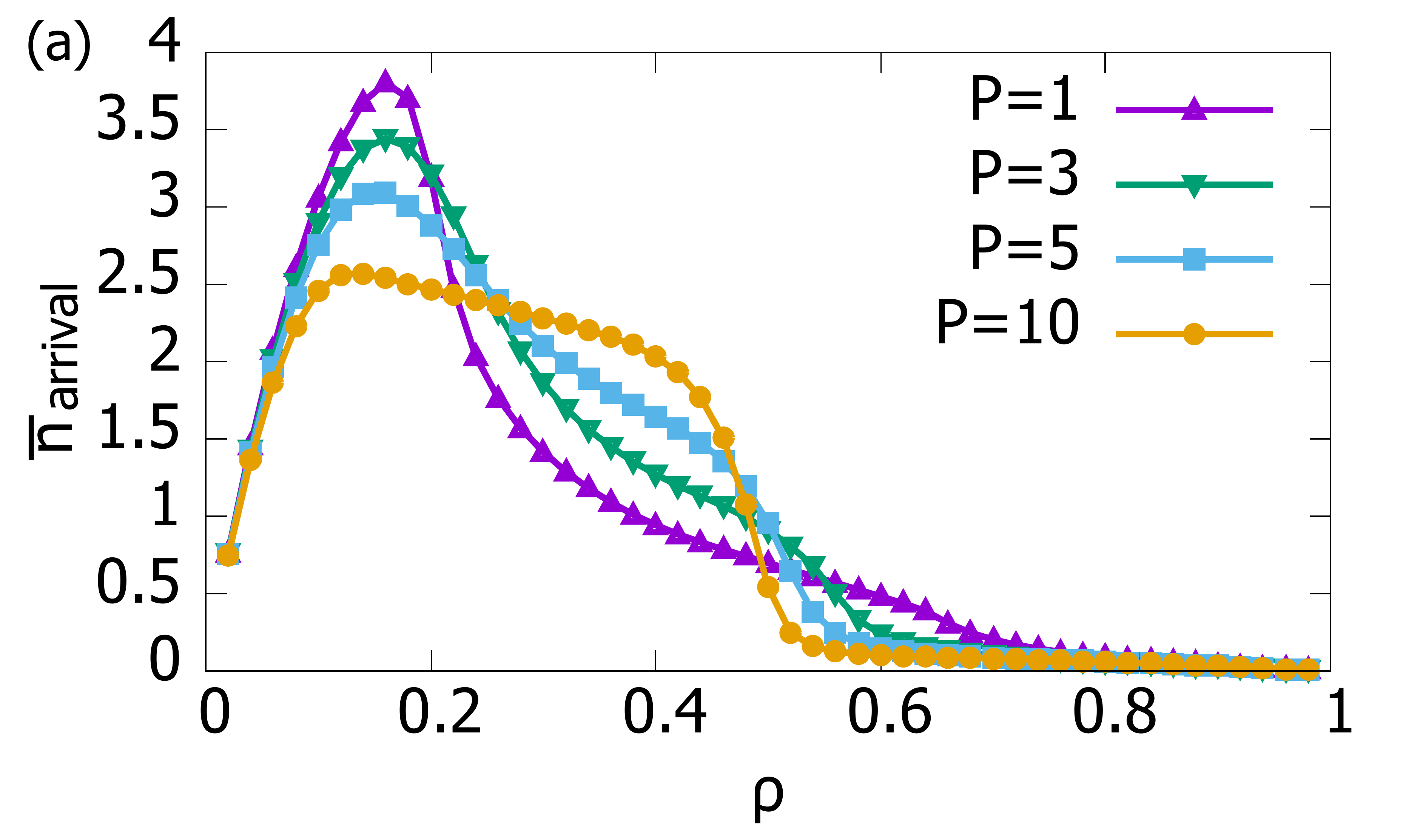}
\end{minipage}\hspace{2pc}%
\begin{minipage}{18pc}
\includegraphics[width=18pc]{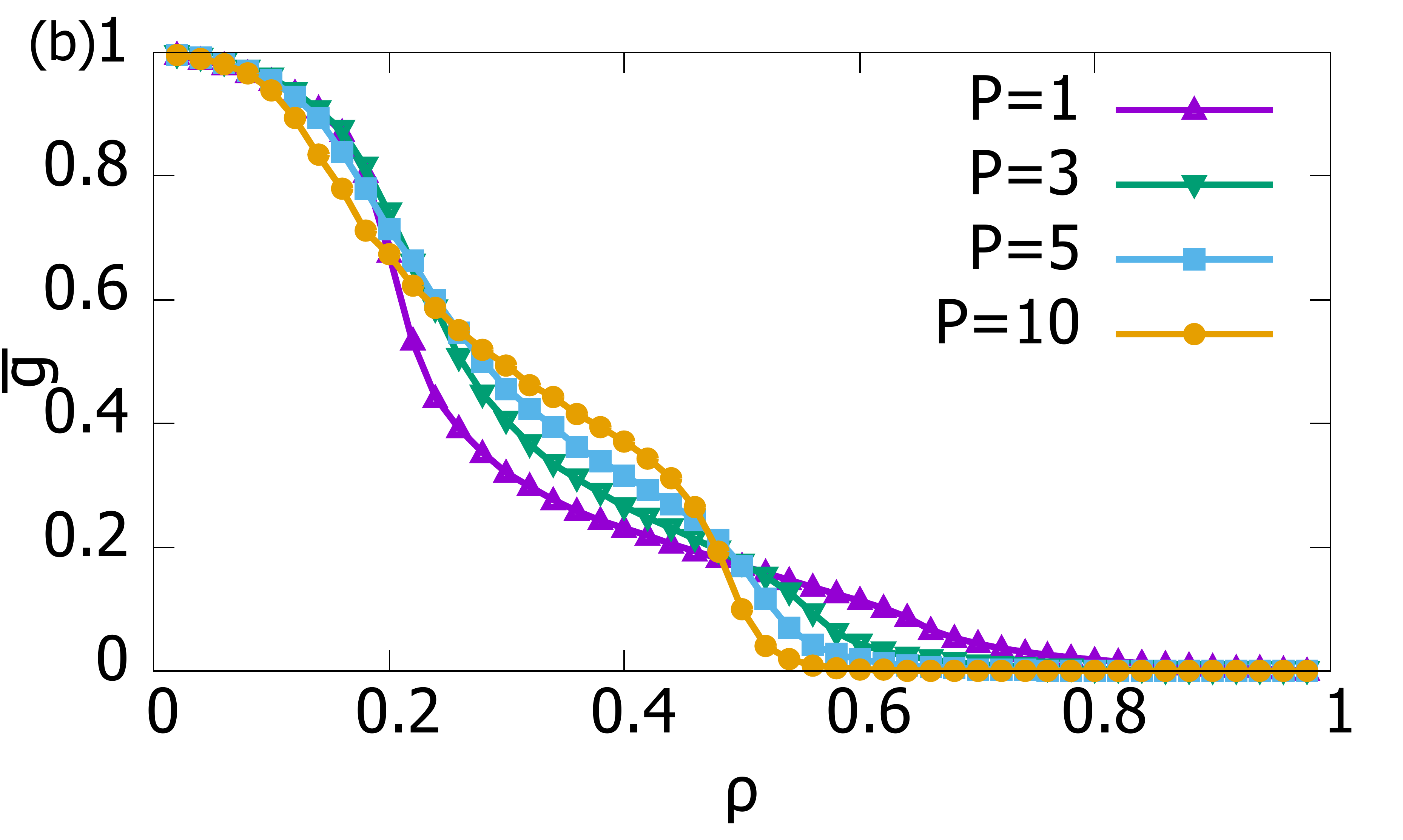}
\end{minipage}
\caption{The simulation results of (a) the average arrival count $\na$ and (b) the average path-greediness $\ng$ as a function of the density of vehicles $\rho$ with $\Dg=0.04$. Results for various values of magnitude change of path-greediness $\Dg$ are obtained on square lattice with $L=20$ and periodic boundary condition.}
\label{diffDP}
\end{figure}

\begin{figure}[h]
\begin{minipage}{18pc}
\includegraphics[width=16pc]{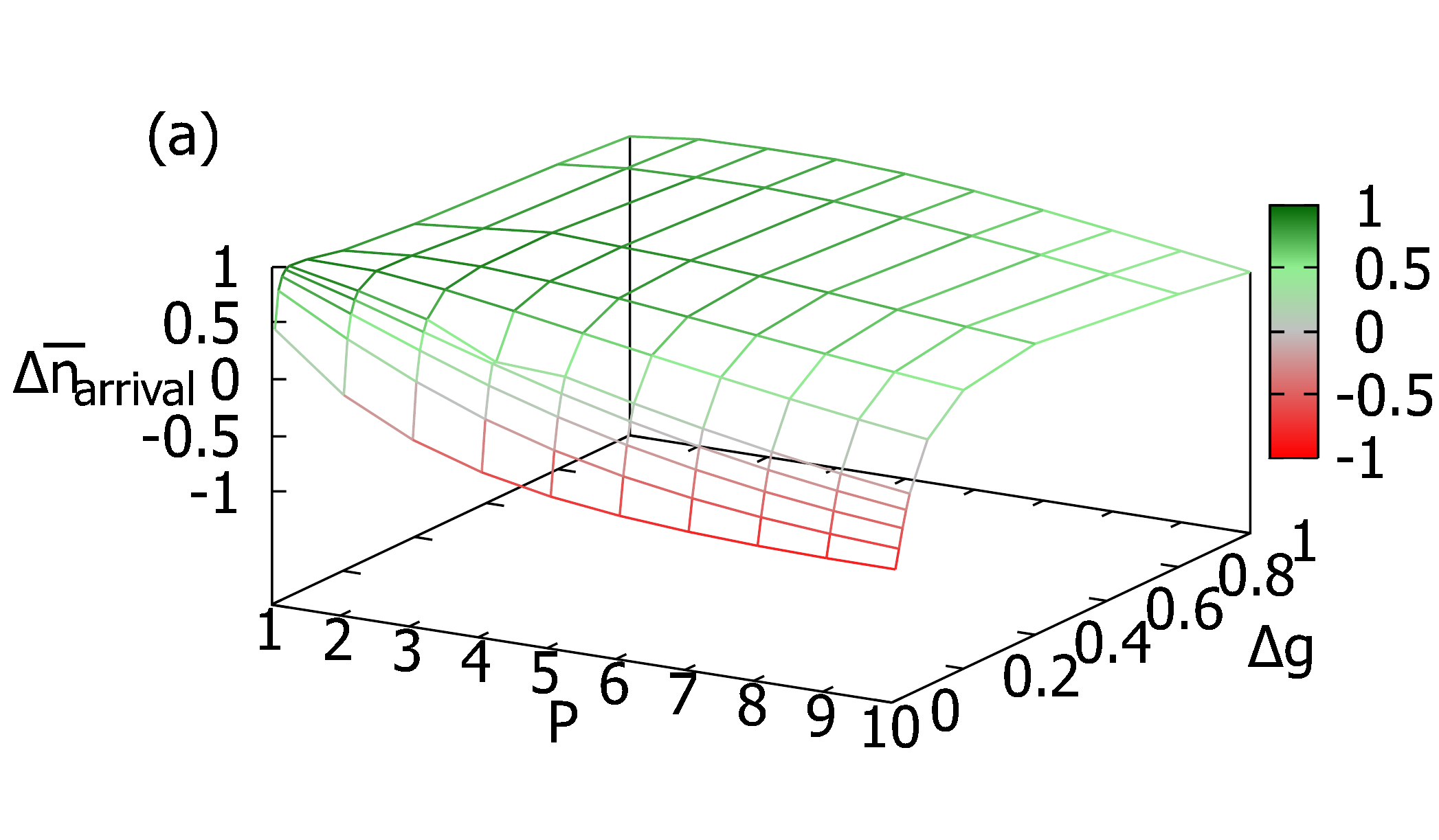}
\end{minipage}\hspace{2pc}%
\begin{minipage}{18pc}
\includegraphics[width=16pc]{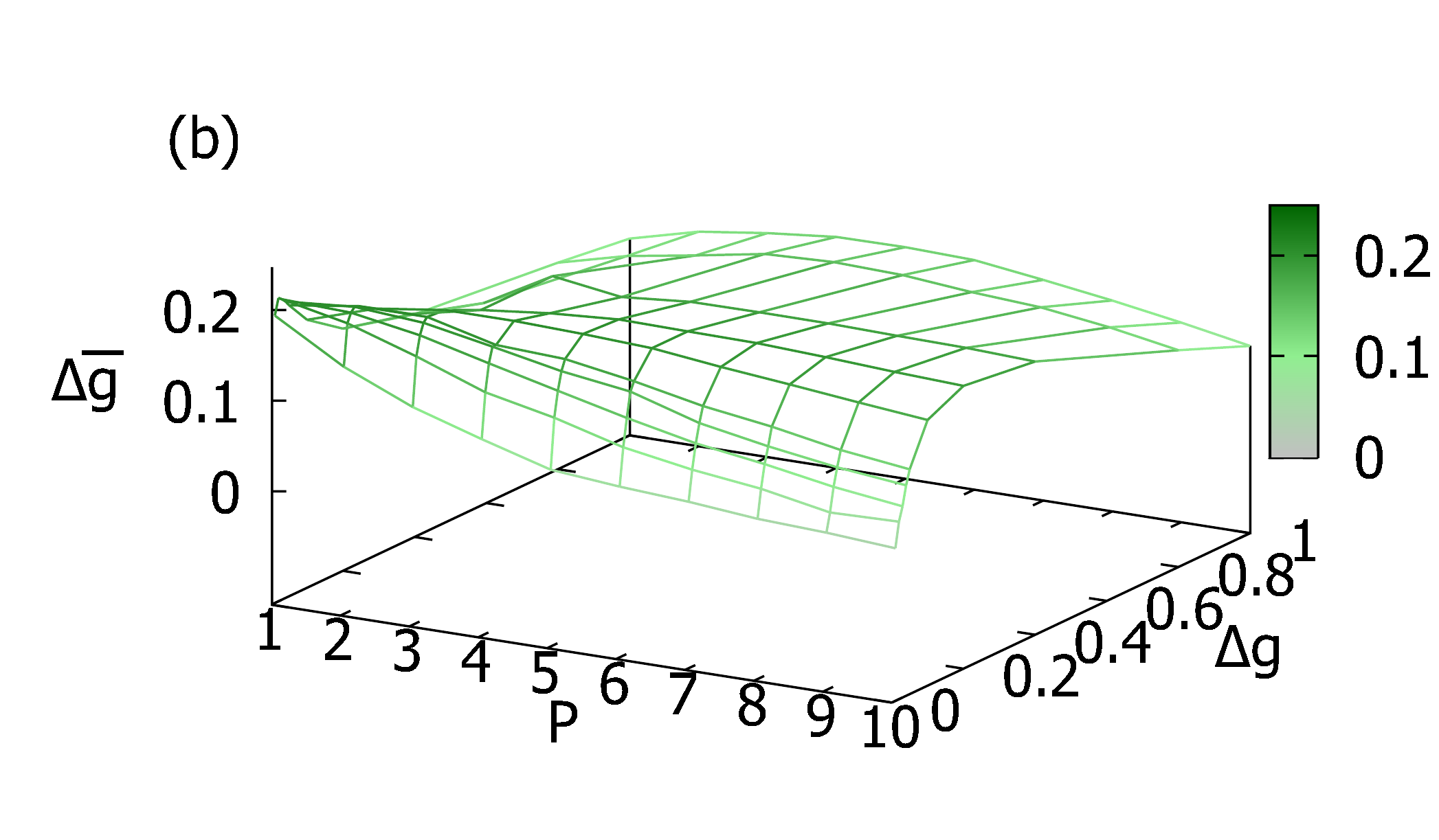}
\end{minipage}
\begin{minipage}{18pc}
\includegraphics[width=16pc]{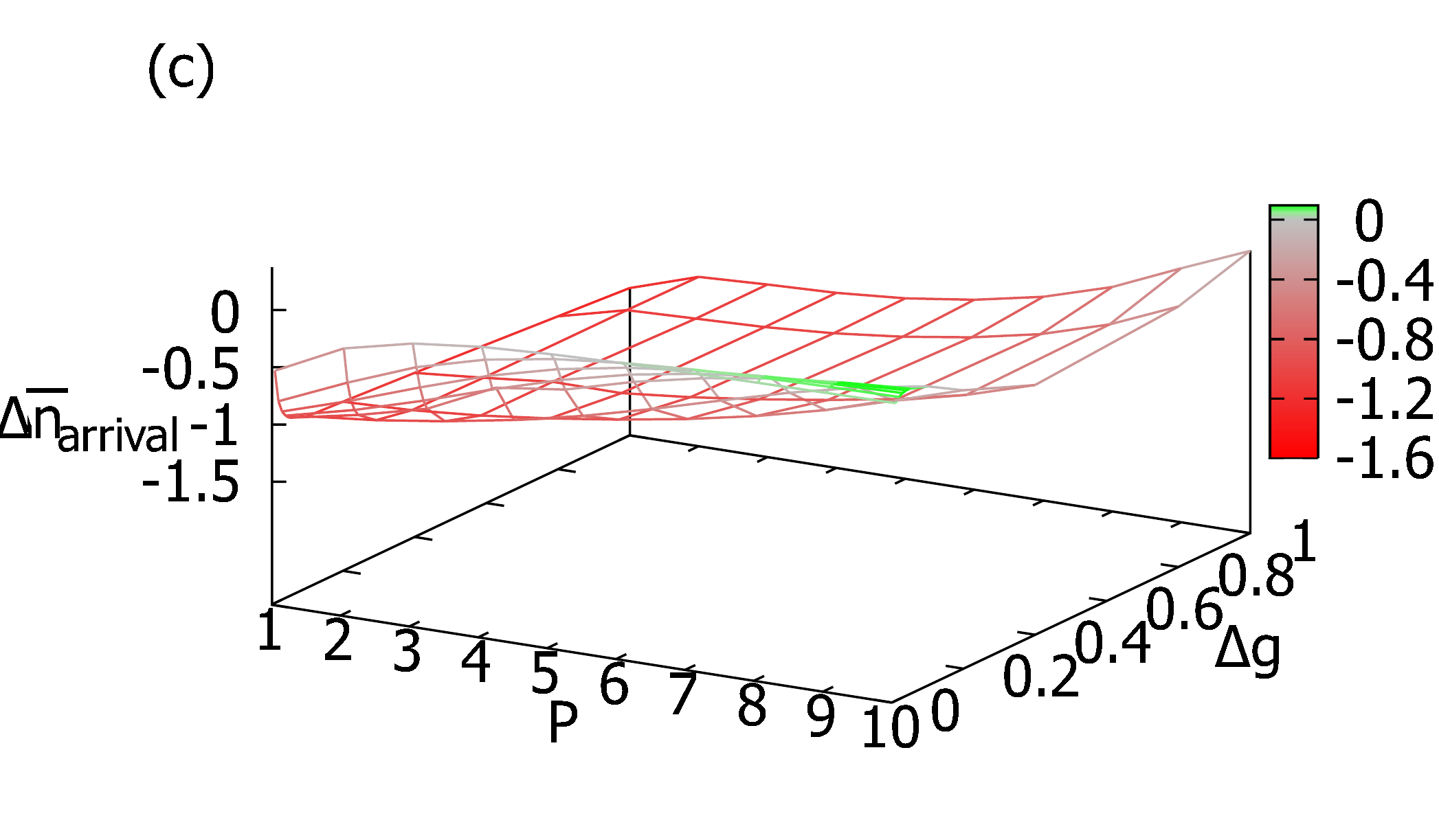}
\end{minipage}\hspace{2pc}%
\begin{minipage}{18pc}
\includegraphics[width=16pc]{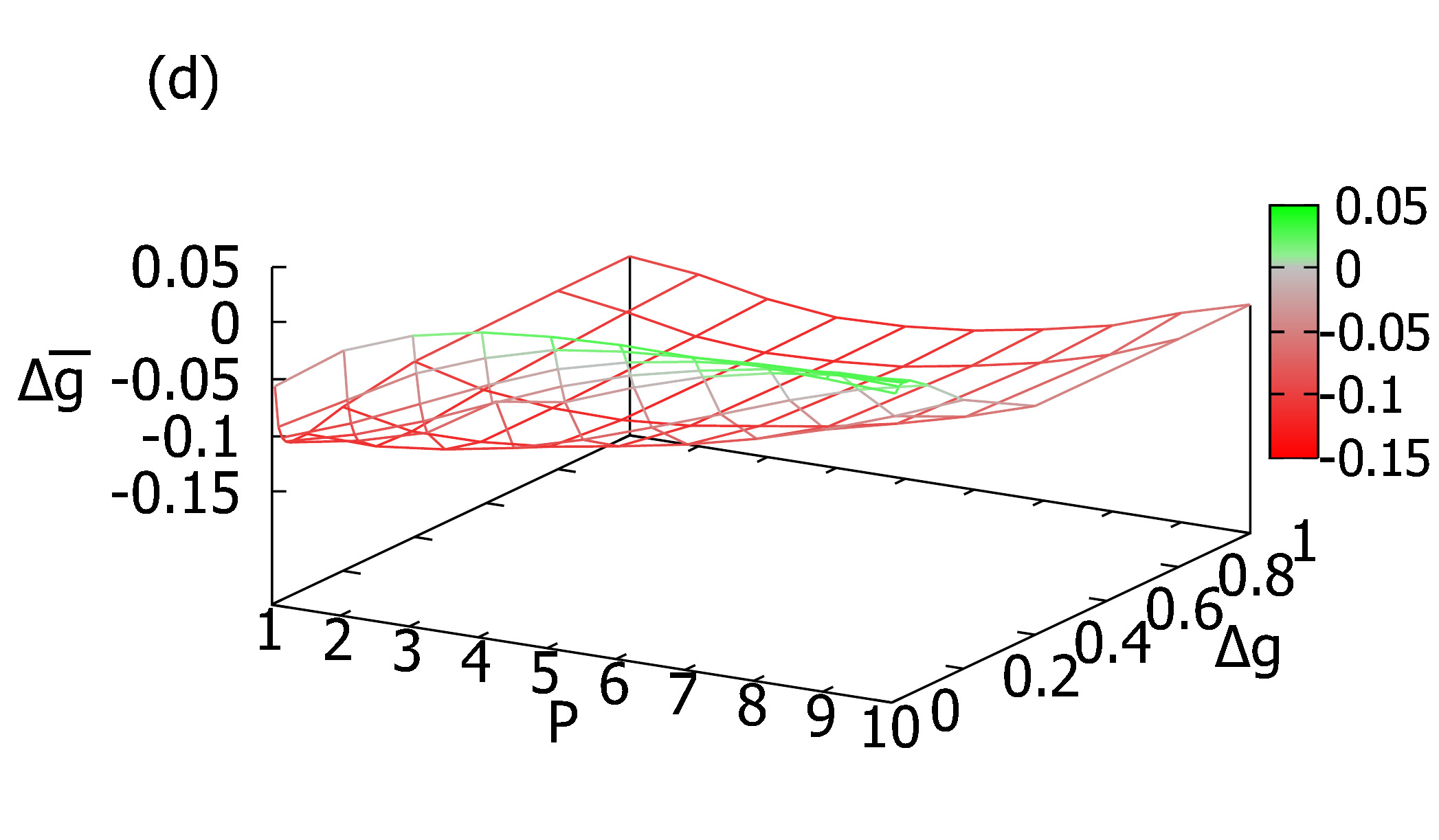}
\end{minipage}
\begin{minipage}{18pc}
\includegraphics[width=16pc]{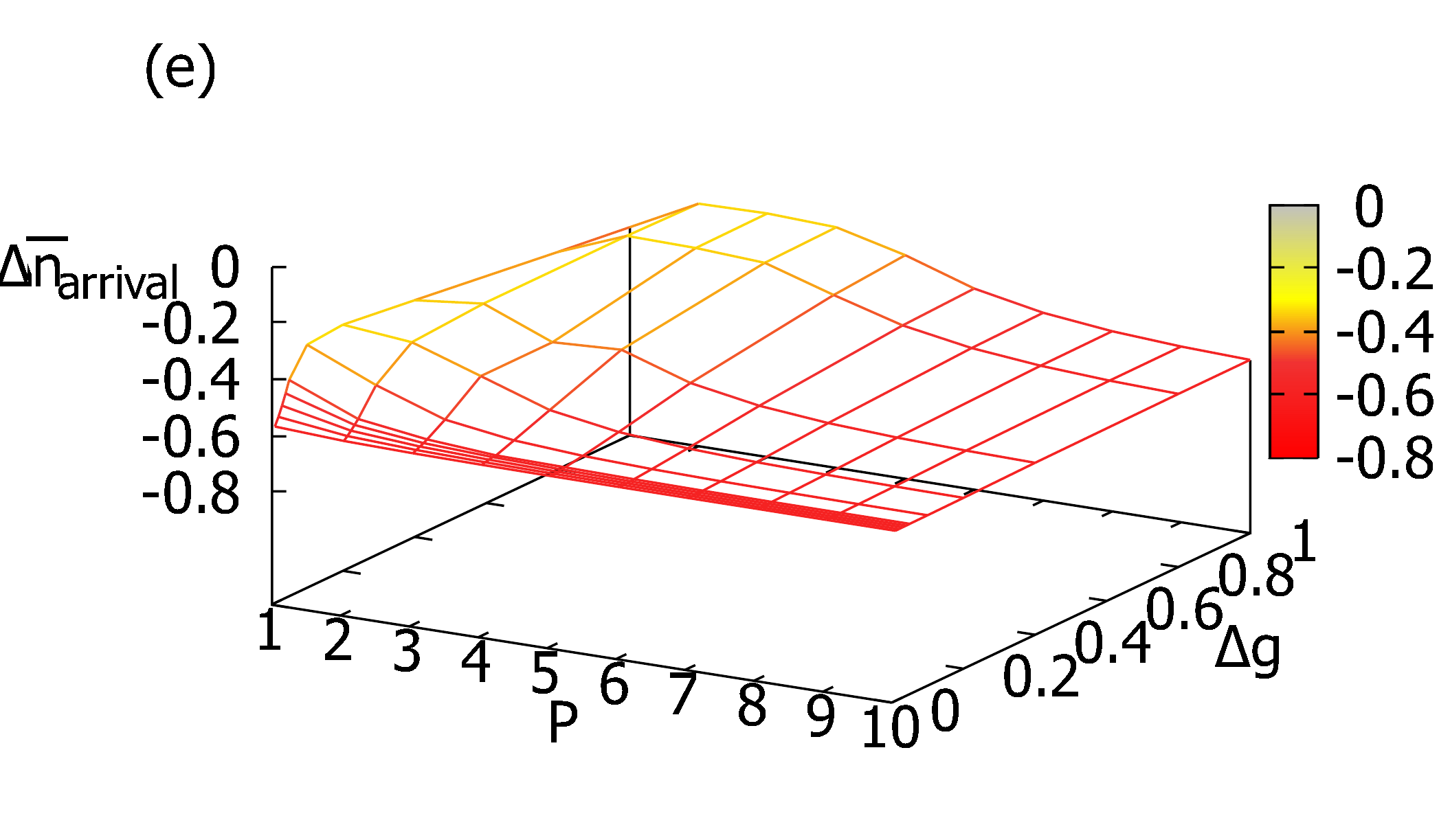}
\end{minipage}\hspace{2pc}%
\begin{minipage}{18pc}
\includegraphics[width=16pc]{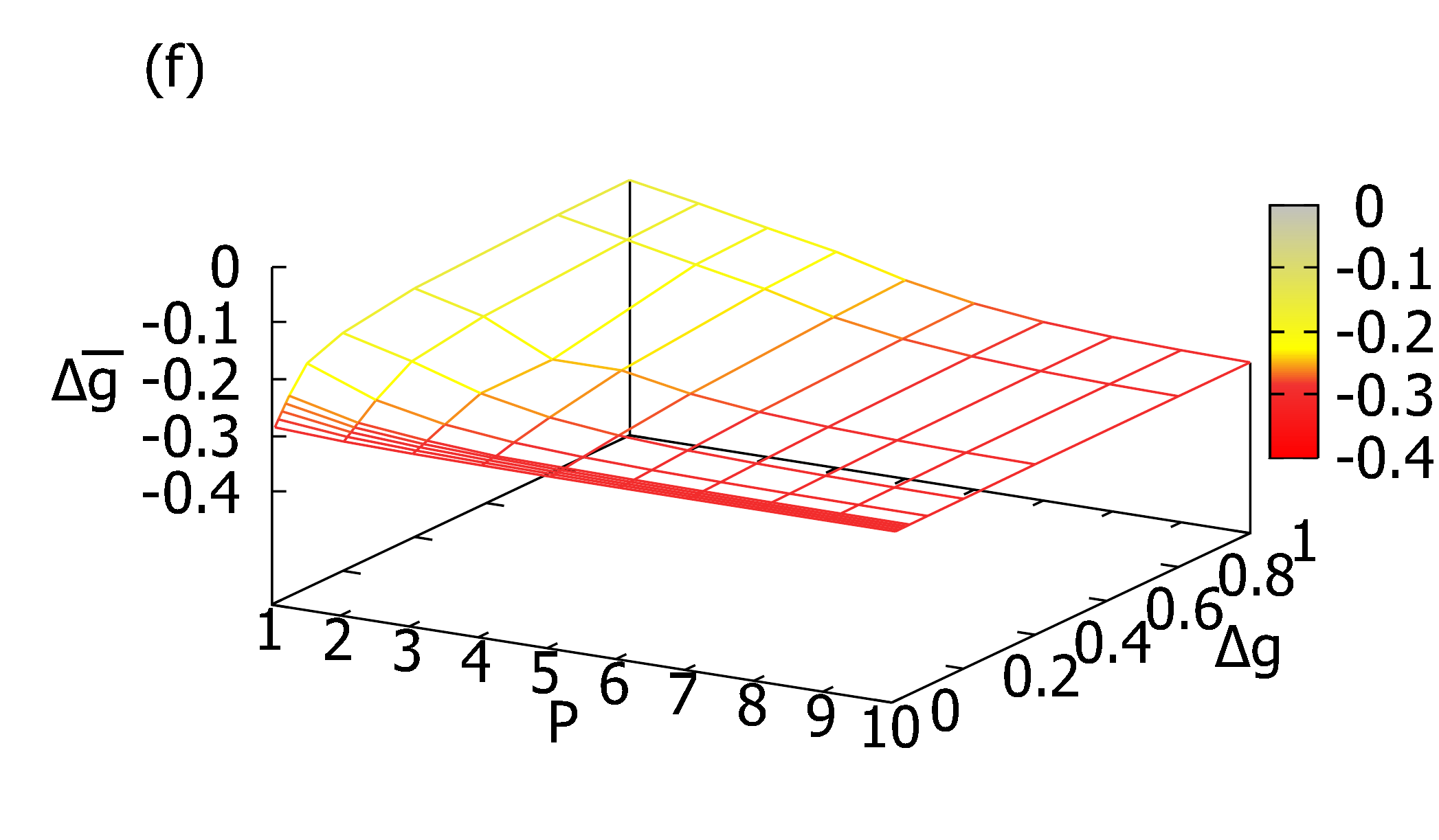}
\end{minipage}
\caption{The simulation results of the difference in arrival count $\Delta \na$ between the adaptive and the controlled cases, as a function of $P$ and $\Dg$ with vehicle density (a) $\rho=0.16$, (c) $\rho=0.4$, and (e) $\rho=0.7$. The corresponding results of the difference in path-greediness $\Delta \ng$ between the adaptive and the controlled cases as a function of $P$ and $\Dg$ with vehicle density (b) $\rho=0.16$, (d) $\rho=0.4$, and (f) $\rho=0.7$.}
\label{deltaArrivalAvgG}
\end{figure}

We then examine the dependence of the system performance on the parameter $P$, i.e. the number of consecutive waiting or movements to trigger an update on path-greediness, while keeping $\Dg$ constant. As shown in \fig{diffDP}(a), the results of the arrival count $\na$ in different cases collapse when $\rho<0.1$. Therefore, in the free-flow state, the dependence of $\na$ on $P$ is insignificant. However, when $\rho$ increases further such that $0.1<\rho<0.4$, the impact of $P$ starts to prevail. As shown in \fig{diffDP}(a), with $0.1<\rho<0.3$,  $\na$ in cases with larger $P$ is lower than that in the cases with smaller $P$. Also, within $0.1<\rho<0.3$, $\na$ is the largest when $P=1$, which is the case that is most responsive to the traffic conditions. It implies that in free-flow state, the value of $g$ should be adjusted quickly (i.e. with a small response time $P$) to prevent serious traffic congestion. 

As shown in \fig{diffDP}(b), at vehicle density $\rho<0.2$, $\ng$ of the case with $P=1$ is higher than that in other cases with $P>1$. It implies that a high response rate can help increase the chances of selecting the optimal path-greediness, and hence the optimal route. However, with $\rho>0.2$, high response rate in updating $g$ is not beneficial, as we can see from their smaller $\na$ and $\ng$. We can see \fig{diffDP}(a) and (b) that when $0.2<\rho<0.56$, both $\na$ and $\ng$ increase with $P$. It implies that keeping the path-greediness for a longer time may be beneficial to the macroscopic performance of the system in the congested state.

We then examine in detail the dependence of model behavior on the parameter $P$ and $\ng$ in the density $\rho=0.16$, $0.4$ and $0.7$, corresponding to the free-flow state, the congested-flow state and the congested state respectively. We first denote the highest number of arrival counts from the \emph{non-adaptive} or \emph{controlled} case  where all vehicles  are characterized by the same value of $g^*_{\rm controlled}$ to be $\nac$. Our previous study~\cite{PhysRevE.100.012311} briefly shows that the adaptive cases outperform the controlled cases in the free-flow state but not in the congested state; nevertheless, only one specific set of values of $P$ and $\Dg$ was compared to the controlled cases in~\cite{PhysRevE.100.012311}, without a comprehensive analysis of their impact on the system which represent the extent of adaptiveness. To further reveal the pros and cons of adaptive routing, we define the difference in arrival counts as well as the resulting average path-greediness between adaptive and controlled cases respectively as

\begin{align}
\Delta \na &= \na - \nac,\\
\Delta g &= \ng - g^*_{\rm controlled}.
\end{align}
We then examine the dependence of $\Delta \na$ and $\Delta g$ on $P$ and $\Dg$. We remark that $\nac$ and $g^*_{\rm controlled}$ are only dependent on density $\rho$, and are independent on $P$ and $\Dg$~\cite{PhysRevE.100.012311}. Our previous results found that in the controlled cases, $\na$ attains its maximum at different $g^*_{\rm controlled}$ depending on $\rho$.

In \fig{deltaArrivalAvgG}(a), we can see that when the system is in the free-flow state at $\rho=0.16$, $\Delta \na>0$ from a majority of parameters $P$ and $\Delta g$, implying that the arrival counts of most adaptive cases outperform those of the controlled cases with optimal $g^*_{\rm controlled}$; the system favors frequent and large updates of path-greediness since positive $\Delta \na$ is found in the region with large $\Delta g$ as well as small $\Delta g$ but with small $P$. As shown in \fig{deltaArrivalAvgG}(b), the resulting self-organized path-greediness in the adaptive cases is larger than $g^*_{\rm controlled}$ for the whole parameter regime, implying that vehicles tend to go to their destinations via the shortest path in the free-flow state.

However, when $\rho$ increases to $0.4$ and the system is in the congested-flow state, as shown in \fig{deltaArrivalAvgG}(c) and (e), $\Delta \na$ from a majority of parameters $P$ and $\Delta g$ become negative, implying that the non-adaptive controlled strategies outperform the adaptive ones. On the contrary to the free-flow state, $\Delta \na$ is only positive with large $P$ and small $\Dg$, implying that the system favors infrequent and small update of path-greediness which is effectively a controlled strategy.  In this case, small adjustment of $g$ over a long period of time in the congested-flow state can benefit arrival counts. Nevertheless, as shown in \fig{deltaArrivalAvgG}(d) and (f), the small region of positive $\Delta \na$ vanishes as $\rho$ further increases to $0.7$ and the system is in the congested state. In this case, the controlled cases always outperform the adaptive ones regardless of $P$ and $\Dg$.

\begin{figure}
\centering
\includegraphics[width=0.8\linewidth]{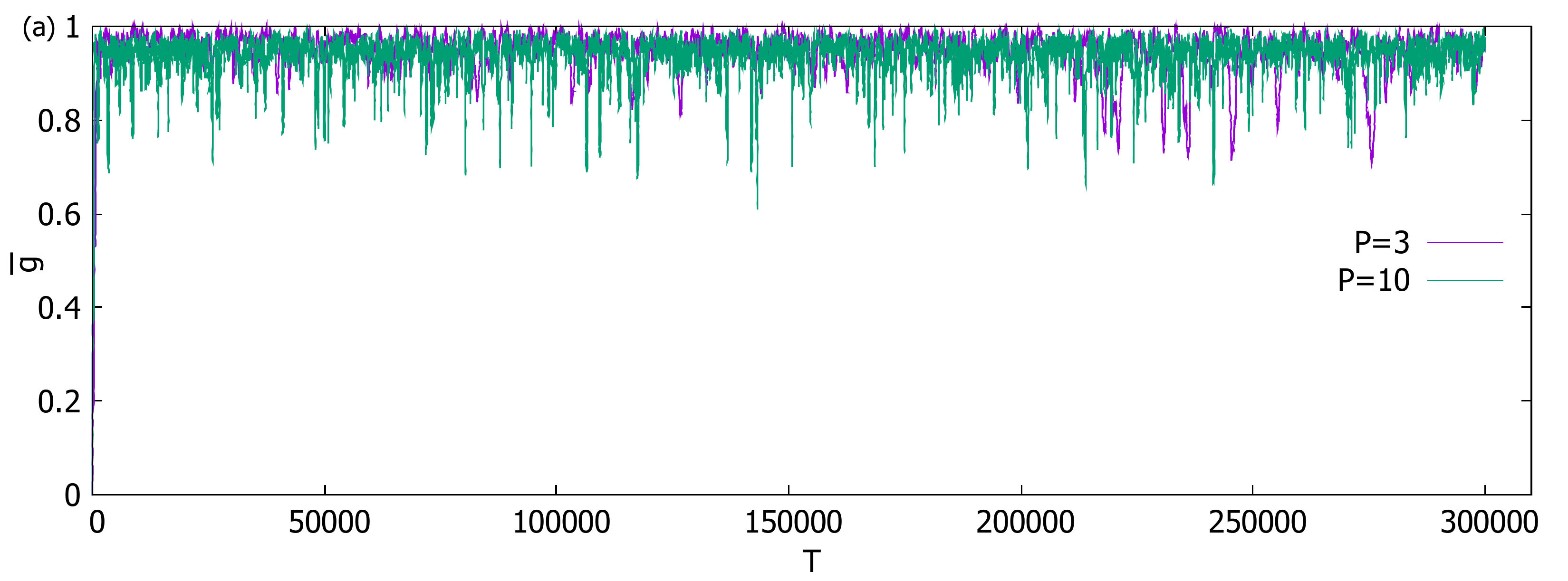}
\includegraphics[width=0.8\linewidth]{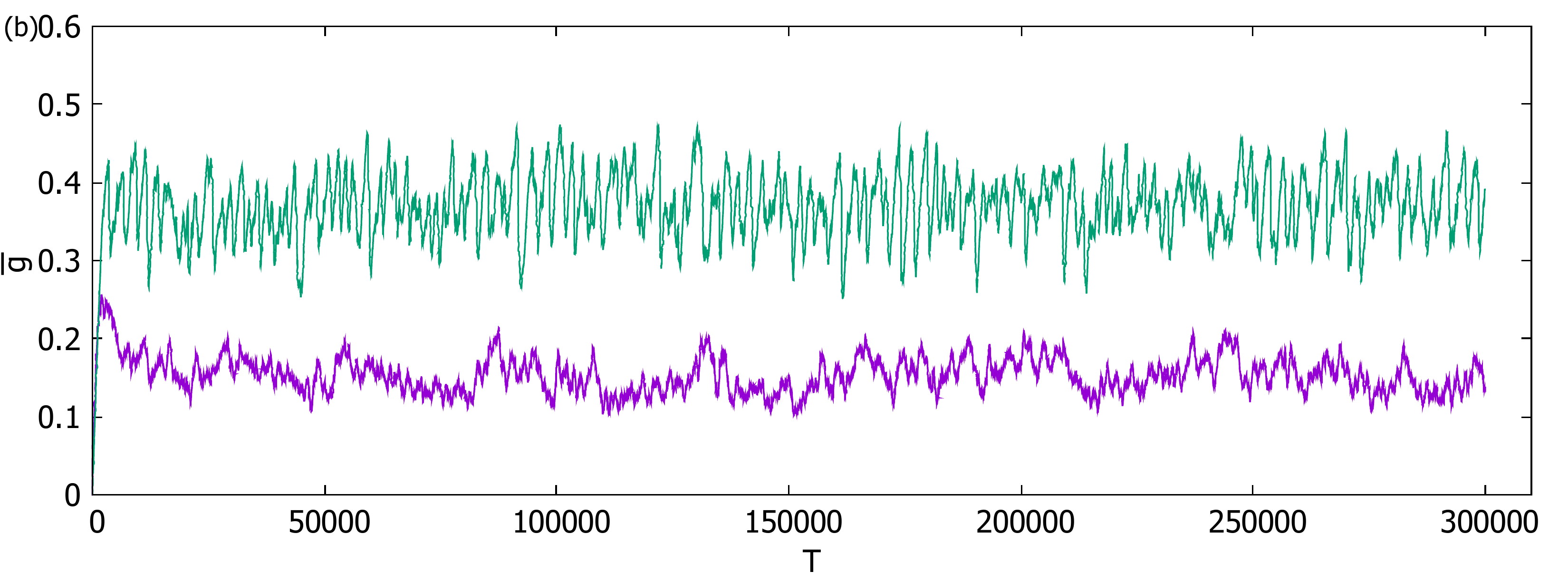}
\includegraphics[width=0.8\linewidth]{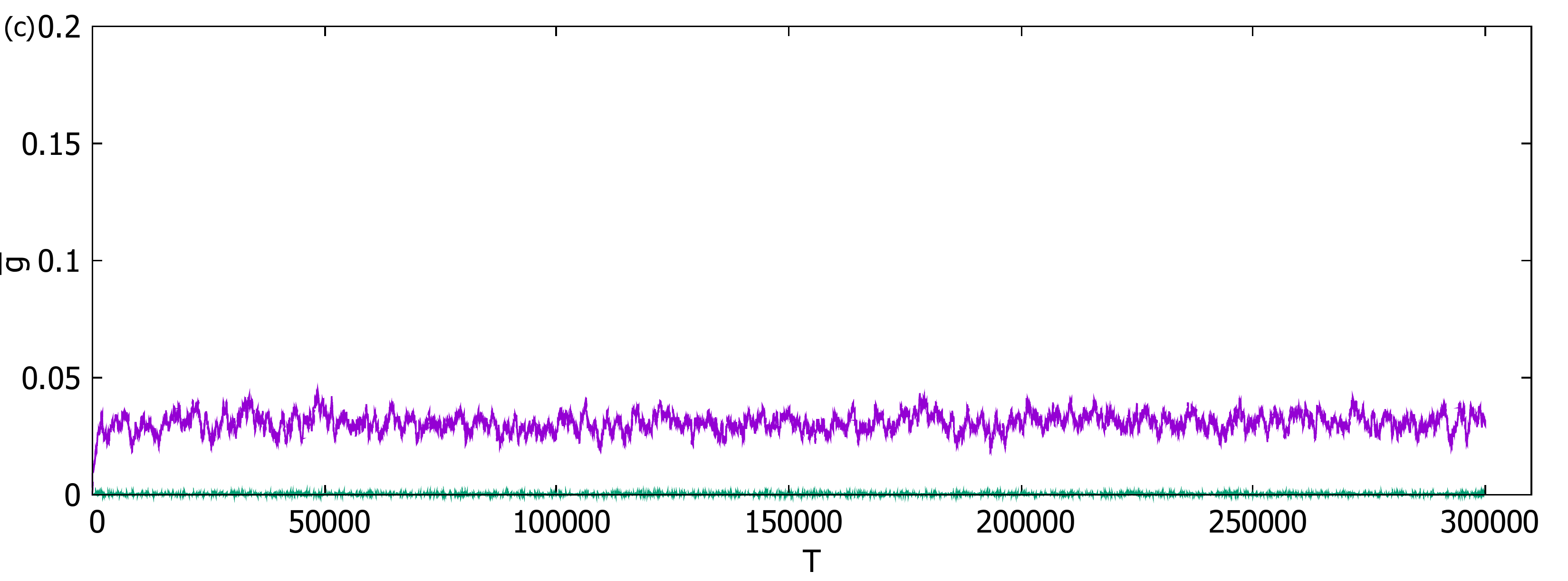}
\caption{The simulation result of average path-greediness $g$ as a function of simulation steps $t$ and various values of density of vehicles (a) $\rho=0.1$, (b) $\rho=0.4$, and (c) $\rho=0.7$ for $L=20$ with periodic boundary conditions and $\Dg=0.04$ in two different samples with $P=3$ and $P=10$.}
\label{timeSeries}
\end{figure}

\begin{figure}[h]
\begin{minipage}{18pc}
\includegraphics[width=16pc]{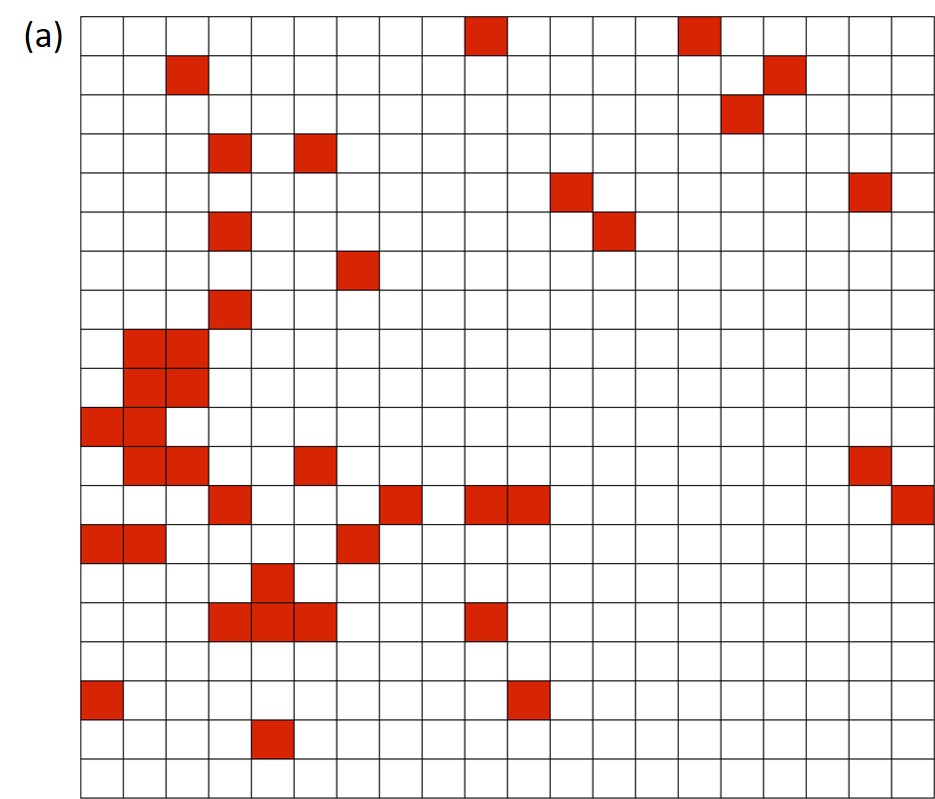}
\end{minipage}\hspace{2pc}%
\begin{minipage}{18pc}
\includegraphics[width=16pc]{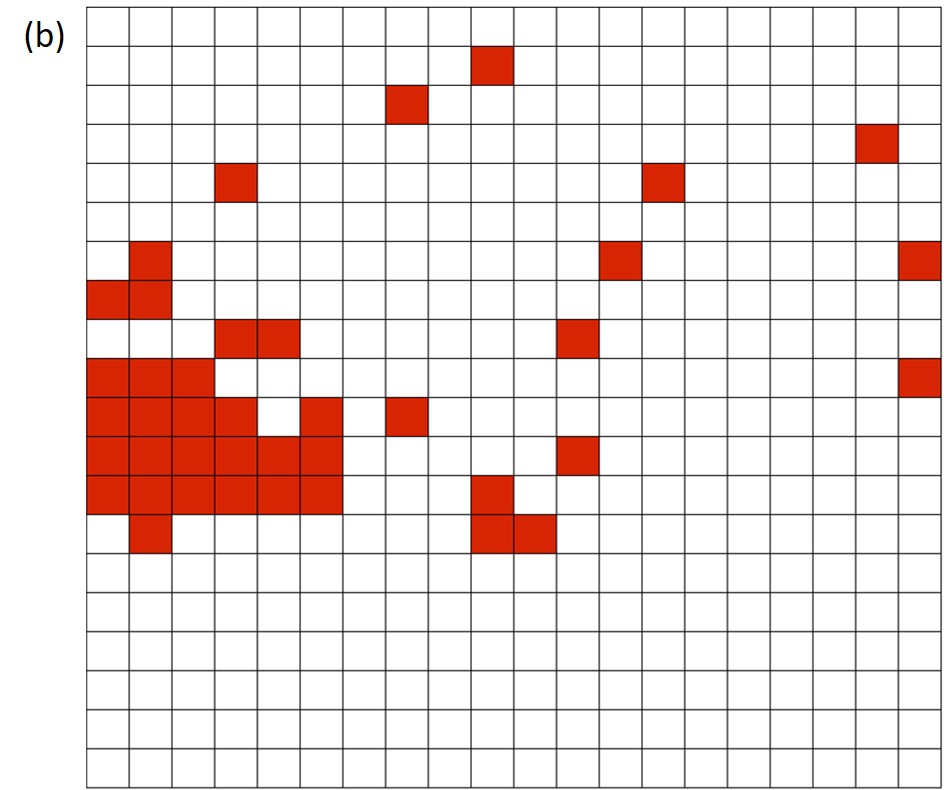}
\end{minipage}
\begin{minipage}{18pc}
\includegraphics[width=16pc]{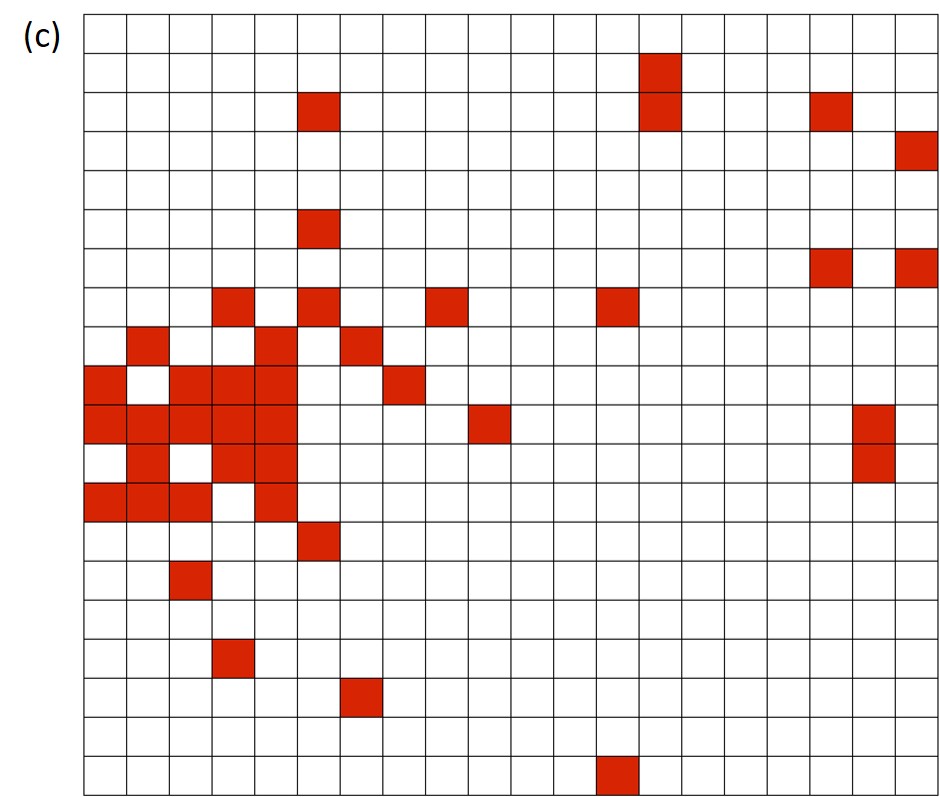}
\end{minipage}\hspace{2pc}%
\begin{minipage}{18pc}
\includegraphics[width=16pc]{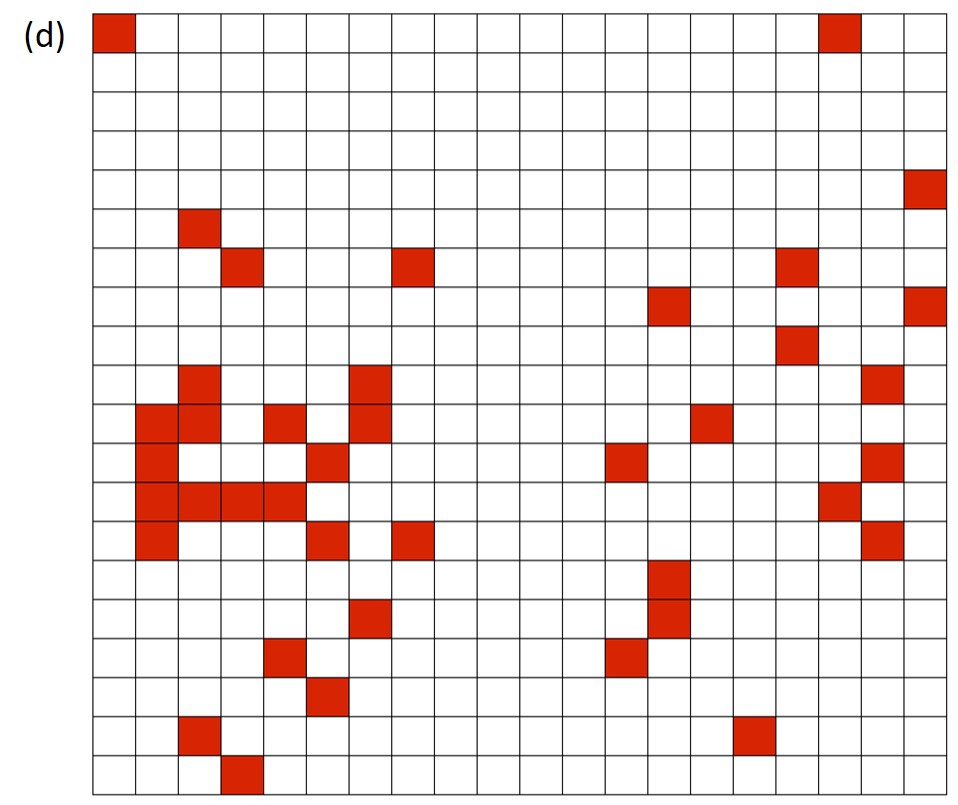}
\end{minipage}
\caption{An example of snapshots of vehicle location on a $20 \times 20$ square lattice with $\rho=0.1$ at different $t$ in the period of temporary congestion in the free-flow state. The sequence is (a), (b), (c) and (d), and the results of $\ng(t)$ are $ 0.981$, $0.642$, $0.629$ and $0.733$ respectively.}
\label{congestedFig}
\end{figure}

To further reveal the pros and cons of adaptive strategies, we examine the system behavior by the time series $\ng(t)$. As $g$ of individual vehicles changes from time to time, $\ng$ changes with time $t$. When $\rho=0.1$, as shown in \fig{timeSeries}(a), $\ng(t)$ usually fluctuates between $0.8$ and $1$, which are high values of path-greediness so that vehicles can mostly travel via the shortest path to their destinations. However, we observe that there are some sudden drops of $\ng(t)$ in the system, i.e. $\ng<0.8$, which implies that traffic congestion temporarily occurs even in the case with a small density $\rho=0.1$, especially with a large $P$ such as $P=10$. For example, in \fig{congestedFig}(a), some vehicles on the left bottom side form a small congested cluster so that $\ng(t)$ starts to decrease. The situation worsens when more vehicles joined the congested cluster, as shown in \fig{congestedFig}(b), causing a sudden drop in $\ng(t)$. The values of $\ng(t)$ continue to decrease and attain the minimum when vehicles start to move away from the cluster, as shown in \fig{congestedFig}(c), since the value of $\git$ of vehicles has decreased to a level that the vehicles route randomly more often and escape from the cluster. Afterwards, as shown in \fig{congestedFig}(d), the vehicles can freely move and $\ng(t)$ increases again and recovers to the level before the congestion. This exemplar temporary congestion lasts around a hundred time steps.

In general in the free-flow state, since temporary congestions cause a drop of $\ng$ over a period of time, the overall $\ng$ with temporary congestions is smaller than the cases without temporary congestions. Moreover, the increase in  vehicle density increases the chances of temporary congestions in the free-flow state, so there is a slight decrease in $\ng$ with the increase in $\rho$ in \fig{diffDg}(b) and \fig{diffDP}(b). We can also see that the number of sudden drops of $\ng$ in the case with $P=3$ is smaller than that in the case with $P=10$, since the response time of vehicles is shorter with $P=3$ and large congested clusters are less likely to emerge. Hence, a more frequency update of $g$ of vehicles can reduce temporary congestions.

When $\rho$ increases beyond $\rho>\rho_c$ to the congested-flow state,  $\ng$ decreases to a low level. As shown in \fig{timeSeries}(b) with $\rho=0.4$, i.e. the congested-flow state, $\ng$ fluctuates between $0.1$ and $0.3$ in the case with $P=3$ whereas between $0.25$ and $0.45$ in the case with $P=10$. In both cases, the fluctuation of $\ng$ is less than that  with $\rho=0.1$ as there is no sudden increase and drop of $\ng$, implying that the traffic condition is stable. As we have shown in \fig{diffDP}, the cases with $P=10$ can maintain a larger value of $\ng$ than the cases with $P=3$.

When $\rho$ further increases to $0.7$, i.e. the congested state, as shown in \fig{timeSeries}(c), the fluctuation of $\ng$ and the value of $\na$ further decreases. In this case, only a small fraction of vehicles have a value of $\git>0$ at each instance, while all the other vehicles have a zero $\git$, and thus the fluctuation in $\ng$ is small, especially with $P=10$. We also observe that in \fig{timeSeries}(c), $\ng$ is never equal to zero in the case with $P=3$, except at $t=0$ when initially all $g_i^0=0$, meaning that vehicles are not totally congested and mutually blocked, and some vehicles can still move in the system and attempt to increase $\git$. On the contrary, $\ng$ is sometimes equal to zero in the case with $P=10$, which means that all vehicles are congested and $P=10$ is too long for vehicles to respond.

In summary, a large magnitude change in path-greediness, i.e. a large $\Dg$, is beneficial in the free-flow state, while a small $\Dg$ is beneficial in the congested state. The results suggest that drivers can choose to be more greedy to travel via the shortest path in the free-flow state, but a more gradual change of driving behavior by drivers en-route to destinations are beneficial in the congested state. On the other hand, a high response rate by drivers is beneficial in the free-flow state but slower responses are preferred in the congested state, which may imply that coordination such as a centrally assigned path-greediness is beneficial in the congested state, consistent with the brief results in~\cite{PhysRevE.100.012311}. In general, adaptive routing is beneficial in the free-flow state but not in the congested state, where the system instead favors non-adaptive controlled routing.

\section{Conclusion}
\label{sec_conclusion}
In this paper, we examine various routing strategies adaptive to the traffic conditions when drivers are en-route to destinations. Our simulation results show that the response rate of drivers in adjusting their routing strategies affect the arrival count of vehicles, especially when the density of vehicles is high. A low response rate to the traffic conditions works better in systems with a high density of vehicles, but on the contrary a high response rate is beneficial for keeping free flows in the networks at low vehicle density. Our results show that, to prevent from under-reacting and over-reacting to the traffic conditions, different strategies should be applied depending on the density of vehicles in the network.

\section*{Acknowledgements}
This work is supported by the Research Grants Council of the Hong Kong Special Administrative Region, China (Projects No. EdUHK ECS 28300215, GRF 18304316, GRF 18301217 and GRF 18301119), the EdUHK FLASS Dean's Research Fund IRS12 2019 04418, ROP14 2019 04396, and EdUHK RDO Internal Research Grant RG67 2018-2019R R4015.

\bibliographystyle{elsarticle-num}
\providecommand{\newblock}{}

\end{document}